\documentclass[aps,prb,twocolumn,showpacs,superscriptaddress,amsmath,amssymb]{revtex4-1}

\usepackage[pdftex]{graphicx}

\begin{document}

%%%%%%%%%%%%%%%%%%%%%%%% TITLE

\title{Anisotropic resistivity of Na$_{1-\delta}$Fe$_{1-x}$Co$_x$As}

\author{N.~Spyrison}

\affiliation{Department of Physics and Astronomy, Iowa State University, Ames, Iowa 50011, USA }
\affiliation{The Ames Laboratory, Ames, Iowa 50011, USA}

\author{M.~A.~Tanatar}
\email[Corresponding author: ]{tanatar@ameslab.gov}
\affiliation{Department of Physics and Astronomy, Iowa State University, Ames, Iowa 50011, USA }
\affiliation{The Ames Laboratory, Ames, Iowa 50011, USA}

\author{Kyuil Cho}
\affiliation{Department of Physics and Astronomy, Iowa State University, Ames, Iowa 50011, USA }
\affiliation{The Ames Laboratory, Ames, Iowa 50011, USA}

\author{Y.~Song}
\affiliation{ Department of Physics and Astronomy, The University of Tennessee, Knoxville, Tennessee 37996-1200, USA}

\author{ Pengcheng Dai }
\affiliation{ Department of Physics and Astronomy, The University of Tennessee, Knoxville, Tennessee 37996-1200, USA}

\author{Chenglin Zhang }

\affiliation{ Department of Physics and Astronomy, The University of Tennessee, Knoxville, Tennessee 37996-1200, USA}

\author{R.~Prozorov}
\email{prozorov@ameslab.gov}
\affiliation{Department of Physics and Astronomy, Iowa State University, Ames, Iowa 50011, USA }
\affiliation{The Ames Laboratory, Ames, Iowa 50011, USA}

\date{17 August 2012}

\begin{abstract}
Temperature-dependent resistivity is studied in single crystals of iron-arsenide superconductor Na$_{1-\delta}$Fe$_{1-x}$Co$_x$As for electrical current directions along, $\rho_a (T)$, and transverse, $\rho_c (T)$, to the Fe-As layers. Doping with Co increases stability of this compound to reaction with the environment and suppresses numerous features in both $\rho_a(T)$ and $\rho _c(T)$ compared to the stoichiometric NaFeAs. Evolution of $\rho_a (T)$ with $x$ follows a universal trend observed in other pnictide superconductors, exhibiting a $T$-linear temperature dependence close to the optimal doping and development of $T^2$ dependence upon further doping. $\rho_c (T)$ in parent compound shows a non - monotonic behavior with a crossover from non-metallic resistivity increase on cooling from room temperature down to $\sim$ 80~K to a metallic decrease below this temperature. Both $\rho_a (T)$ and  $\rho_c (T)$ show several correlated crossover - like features at $T>$ 80~K. Despite a general trend towards more metallic behavior of inter - plane resistivity in Co-doped samples, the temperature of the crossover from insulating to metallic behavior (80 K) does not change much with doping.
\end{abstract}


%Metals, transport processes in, 72.15.-v

%Superconducting materials 74.70.Dd Ternary, quaternary, and multinary compounds (including Chevrel phases, borocarbides, etc.)

%74.25.Jb Electronic structure

\maketitle

%%%%%%%%%%%%%%%%%%%%%%%%%%%% INTRODUCTION

%%%%1) Introduction
%Oxypnictide superconductors, discovery, properties

\section{Introduction}

Structurally, iron based superconductors are layered materials, in which FeAs (or iron chalcogenide) layer is the main building block for a variety of compounds \cite{paglione,Johnston,stewart}. Since the dominant contribution to the density of states at the Fermi level comes from the iron 3d orbitals, one can expect a significant electronic anisotropy of the compounds revealed in the in-plane and out-of-plane transport. Contrary to this expectation, the most studied families of iron arsenides, those based on BaFe$_2$As$_2$, have rather low anisotropy ratio $\gamma_{\rho}$ $\equiv$ $\rho_c $/$\rho_a$ $\sim$4 at $T_c$ \cite{anisotropy}. In transition metal-doped Ba(Fe$_{1-x}M_x$)$_2$As$_2$ ($M$=Co, Ni, Rh, Pd, BaT122 in the following), $\rho_c (T)$ also shows a very different temperature dependence compared with $\rho_a (T)$, revealing a broad crossover from non-metallic to metallic temperature dependence assigned in our systematic doping studies to the formation of a pseudogap \cite{anisotropy,anisotropy2,pseudogap,pseudogap2}.

Another interesting feature of iron arsenides that distinguishes them from the copper oxide based (cuprate) superconductors \cite{cupratesresistivity} is a strong variation of the functional form of temperature-dependent resistivity for various types and levels of dopings. The general trend of $\rho_a (T)$ evolution is the presence of a $T-$linear region immediately above $T_c$ for optimally doped compositions \cite{NDL,WenK,Kasahara,NaFeAs}. At higher temperatures this $T$-linear behavior, for example in hole-doped Ba$_{1-x}$K$_x$Fe$_2$As$_2$ and in self-doped Na$_{1-\delta}$FeAs, is terminated by the pseudogap \cite{TanatarK,NaFeAs}.

Systematic studies of the temperature-dependent electrical resistivity are very important for the general understanding of superconductivity in this family of materials. Scattering in the normal state in the vicinity of the magnetic quantum critical point leads to the characteristic $T-$linear temperature dependence of $\rho_a (T)$, which then evolves towards Fermi-liquid $T^2-$  behavior with doping (see Ref.~[\onlinecite{Louis}] for a review). Deviations from this general behavior provide an insight into electronic and magnetic correlations \cite{Fernandes}, in particular, into the mechanism of nematic state formation \cite{Fisher,detwinning,ECBK}.

In this article we report the systematics of doping-evolution of in-plane and inter-plane resistivity of electron-doped Na$_{1-\delta}$Fe$_{1-x}$Co$_x$As. This compound shows a ``dome - like'' phase diagram which is very similar to BaCo122 \cite{phaseDNaFeAs,phaseD2NaFeAs,NaFeAs}. As such, this study brings additional insight into the scattering and correlation phenomena of the iron - based superconductors.

\section{Experimental}

AC magnetic characterization of the samples was performed with a tunnel-diode resonator, (TDR)\cite{Vandegrift,Prozorov2000}. Briefly, TDR is a self-oscillating $LC$ tank circuit powered by a properly biased tunnel diode. The sample is mounted with Apiezon N-grease on a sapphire rod and is inserted in the inductor (coil). The sample temperature is controlled independent of the resonant circuit, which is actively stabilized at the constant temperature. The measured frequency shift is proportional to the differential magnetic susceptibility of the sample \cite{Prozorov2000}. In this work, for quick mounting and measurement protocols we used a simplified version of the TDR susceptometer (a ``dipper''), which is inserted directly into a transport $^4$He dewar and gives very quick turn-around measurement time of typically 30~minutes per sample. The trade-off of this quick measurement protocol is reduced stability and higher temperature-dependent background as compared to our high-stability $^3$He and dilution refrigerator versions of the TDR susceptometer. Nevertheless, the ``dipper'' is perfectly suitable to study magnetic signature of the superconducting transition.

Single crystals of Na(Fe$_{1-x}$Co$_x$)As with $x =$ 0, 0.025, 0.05, 0.08, and 0.10 were synthesized by sealing a mixture of Na, Fe, As, and Co together in Ta tubes and heating it to 950 $^{\circ}$C, followed by 5 $^{\circ}$C/hour cooling down to 900 $^{\circ}$C \cite{He2010PRL}. The $x$ in samples was defined as the nominal ratio, which gives some variation with electron-probe microanalysis values \cite{XHChenNaFeAs,ShiyanNaFeAs}. The samples were stored and transported in sealed containers filled with inert gas.

Sample preparation was done quickly in air within about 5 minutes to minimize uncontrolled environmental exposure which can induce an increase of $T_c$ \cite{Todorov2010CM,NaFeAs}. We started sample preparation by cleaving slabs from the inner part of the crystals with typical a thickness of 50 to 100 $\mu$m. The slabs had shiny cleavage surfaces and were further cut into smaller pieces for TDR (typically 0.5$\times$0.5mm$^2$) and resistivity measurements. Cleaved internal parts of single crystals did not show any visible reaction with air and turned out to be relatively stable, contrary to crystals with the residue of NaAs flux, which aggressively reacts with air and moisture. After preparation, samples were promptly measured and immediately stored after measurements in inert and dry environment. After each dipper run samples were washed with toluene to remove remaining N-grease in order to control the air exposure.

Samples for in-plane resistivity measurements had typical dimensions of (1-2)$\times$0.5$\times$(0.02-0.1) mm$^3$. All sample dimensions were measured with an optical microscope with an accuracy of about 10\%. Sample resistivity at room temperature, $\rho (300K)$, was in the range 400 to 500 $\mu \Omega cm$ for all compositions studied. This value is obtained on a bigger array of samples than in our previous study \cite{NaFeAs} and is somewhat higher. It is also somewhat higher than values found in electron- \cite{Alloul,pseudogap} and hole-doped \cite{WenK} Ba122 compounds, typically 300 $\mu \Omega cm$ or less. We do not have sufficient array of data to obtain lower error bars needed to resolve the doping evolution of $\rho (300K)$, if any exists. Contacts for four-probe resistivity measurements were made by soldering 50 $\mu$m silver wires with ultra-pure Sn solder, as described in Ref.~[\onlinecite{SUST}]. Resistivity measurements were performed in {$\it Quantum Design$} PPMS, providing magnetic fields up to 9~T. For measurements of the upper critical field, $H_{c2}$, samples were glued to the side of a plastic block with $ab$ plane of the sample oriented to be either parallel or perpendicular to the direction of magnetic field (with an accuracy of about 1$^\circ$).

Inter-plane resistivity measurements were done using the two-probe technique, relying on very low contact resistance of soldered contacts, typically in the 10 $\mu \Omega$ range. The top and bottom surfaces of the $ab$-plane (typically 0.5$\times$0.5mm$^2$ area) of the samples were covered with Sn solder forming a capacitor-like structure.
A four-probe scheme was used to measure a sample with contacts, giving a sum of series connected sample, $R_s$, and contact resistance, $R_c$ resistances. Since $R_s \gg R_c$, contact resistance represents a minor portion, on the order of 1-5\% on the total resistance. This can be directly seen for our samples for temperatures below the superconducting $T_c$, where $R_s =$0 and the measured resistance represents $R_c$ \cite{anisotropy,SUST,vortex}. Further details of the measurement procedure can be found in Refs.~\onlinecite{anisotropy,anisotropy2,pseudogap}.

The drawback of the measurement of samples with $c \ll a$ is that any structural and chemical inhomogeneity along the $c-$axis, a very common problem in soft and micaceous samples of iron arsenide superconductors \cite{NiNiCo,anisotropy,Bobkowski}, not only increases sample resistance, but admixes in-plane component due to the redistribution of the current. One way to ascertain correctness of the $\rho _c$ measurements, is to rely on measurements with the lowest resistivity values. Typically the best results were obtained on the thinnest slabs. To get reliable results we performed measurements of $\rho_c$ on at least 5 samples of each batch. In all cases we obtained qualitatively similar temperature dependences of the normalized electrical resistivity, $\rho _c (T)/\rho _c (300K)$. The resistivity value at room temperature, $\rho_c (300K)$, however, showed a notable scatting and was typically in the range 2000 to 3000 $\mu \Omega$~cm at room temperature.

We have shown previously that reaction with air strongly affects the value of $\rho _a (300K)$ due to the development of cracks \cite{NaFeAs}. Cracks grossly effect the internal sample connectivity and, hence, homogeneous current distribution, thus, making inter-plane resistivity measurements of environmentally exposed samples impossible.

\section{Results}

\subsection{Environmental stability}

%%%%%%Figure1 TDR air time
\begin{figure}[tbh]%
\includegraphics[width=8cm]{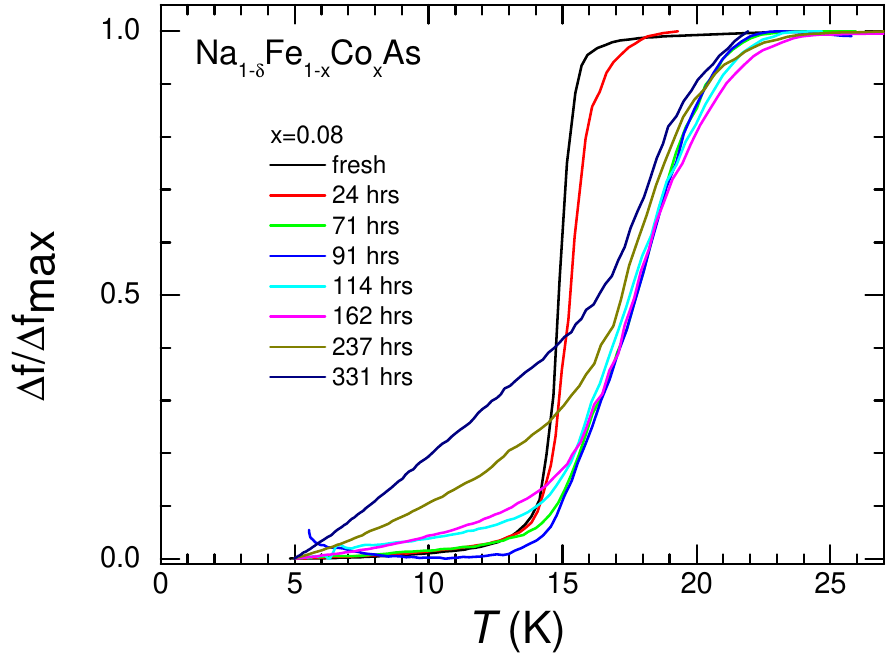}%
\caption{(Color online) Evolution of the frequency shift signal in the dipper TDR experiment on increasing time of air-exposure treatment in crystals of slightly overdoped Na$_{1- \delta}$Fe$_{1-x}$Co$_x$As $x$=0.08. Similar to parent compound \cite{NaFeAs}, $T_c$ of the sample increases with exposure time to a maximum and then decreases. Similar effects are observed for other doping levels, $x$. }%
\label{TDRairtime}%
\end{figure}

%%%%%%Figure2 TDR air
\begin{figure}[tbh]%
\includegraphics[width=8cm]{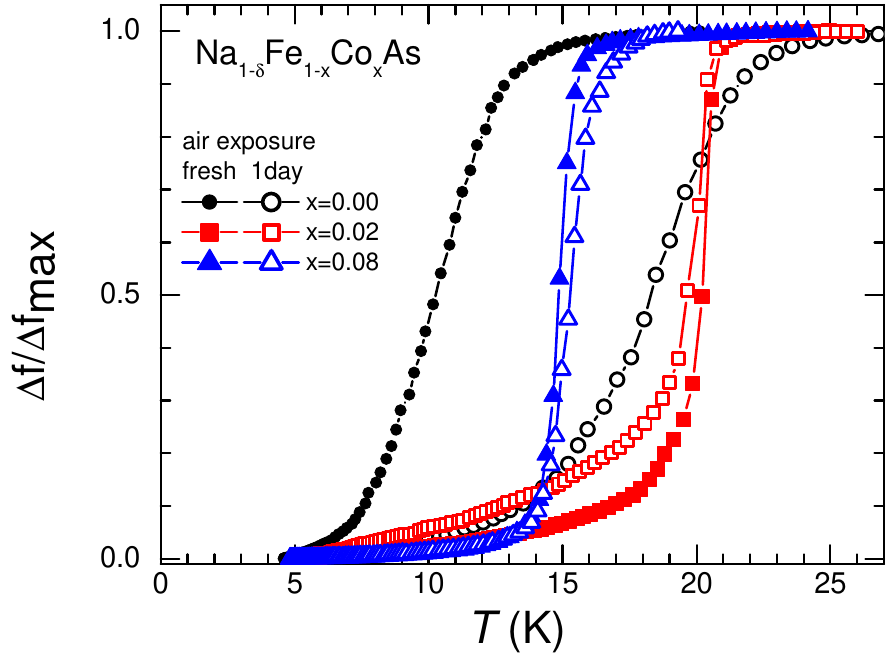}%
\caption{(Color online) Doping evolution of the frequency shift signal in the dipper TDR experiment during fixed time air-exposure treatment of NaFe$_{1-x}$Co$_x$As for 24 hours. As can be seen, $T_c$ increase from 12~K to 22~K, characteristic of parent NaFeAs samples, is strongly suppressed with Co doping.  }%
\label{TDRair}%
\end{figure}

%%%%%%Figure3 TDR air summary
\begin{figure}[tbh]%
\includegraphics[width=8cm]{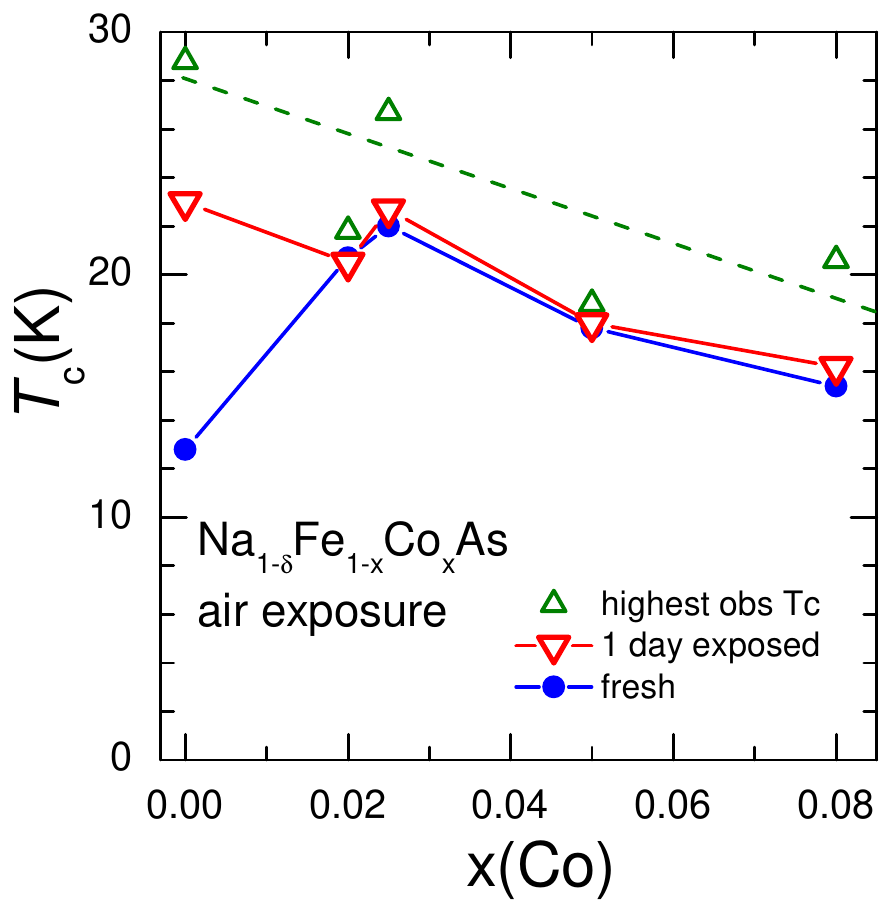}%
\caption{(Color online) Doping evolution of the superconducting $T_c$ in the dipper TDR experiment during air-exposure treatment of NaFe$_{1-x}$Co$_x$As. Blue solid dots show $T_c$ of fresh samples, red down-triangles $T_c$ of the samples exposed to air for one day. Green up-triangles show maximum onset $T_c$ obtained in these experiments. }%
\label{TDRairsummary}%
\end{figure}

The $T_c$ of the parent Na$_{1-\delta}$FeAs increases significantly upon exposure to air \cite{Chu,NaFeAs}, water \cite{Todorov2010CM} and Apiezon N-grease \cite{NaFeAs}. Here, we study how the sensitivity of $T_c$ to exposure changes with Co- doping. In Fig.~\ref{TDRairtime} we show the evolution of the TDR signal in slightly overdoped, $x$=0.08 (fresh sample  $T_c$=16~K), samples on exposure to air. Similar to the parent compound \cite{NaFeAs}, for all compositions irrespective of their $x$, the $T_c$ of the samples increases initially upon air exposure and then decreases with prolonged exposure. The doping-variation of TDR signal during fixed time, one day air exposure, is summarized in Fig.~\ref{TDRair}. Variation of fresh sample $T_c$  and highest achieved $T_c$ during air exposure and fixed - time of one day exposure as a function of $x$ are summarized in Fig.~\ref{TDRairsummary}.

One day exposure of a sample to air does not lead to a visual appearance of reaction products. Thus, at least at this initial stage, there is no reason to assume transformation of NaFeAs into NaFe$_2$As$_2$, the final product of reaction with water \cite{Todorov2010CM}, formed after one a month exposure, which most likely shows up in Fig.~\ref{TDRairtime} as a new shoulder in the temperature dependent frequency shift at about 12~K for samples exposed for about two weeks.

In parent Na$_{1-\delta}$FeAs the environmental reaction is caused by the variation of Na content in the samples, $\delta$, due to oxidative deintercalation \cite{Todorov2010CM}. It is natural to expect similar effect in the Co - doped NaFe$_{1-x}$Co$_x$As. However, the puzzling observation is that $T_c$ increases for both environmental reaction in pure Na$_{1-\delta}$FeAs (presumably hole - like doping) and electron Co - doping. As such, it is not clear if carrier type and density change is the main effect involved. We note that detailed study of the effect of Li deficiency in a closely related LiFeAs superconductor found suppression of $T_c$, but virtually no change in the normal - state properties \cite{LiDeficiency}. If Na deficiency leads to the formation of Na vacancies, this should lead to hole doping and, thus, move the dome on the doping phase diagram in an opposite way to electron Co-doping. Further studies are required to understand what type of doping is induced by the loss of Na and what types of defects are formed.

At a first glance, the different rate of $T_c$ variation in Fig.~\ref{TDRairsummary} can be attributed to the different sensitivity of $T_c (x)$ in different parts of the phase diagram - being smallest at the flat optimal doping region. Indeed, the slope of $T_c (x)$ changes from very high for the parent compound to negligible in samples close to optimal doping. However, we find a rise in $T_c$ after environmental reaction even in over-doped samples. This contradicts the simple relation of the rate of $T_c$ change to be determined by a position on the phase diagram. Our observations rather support previous observation \cite{XHChenNaFeAs} that Co doping increases the stability of the samples.

\subsection{Resistivity measurements}

%%%%%%Figure4 In-plane Resistivity
\begin{figure}[tbh]%
\includegraphics[width=8cm]{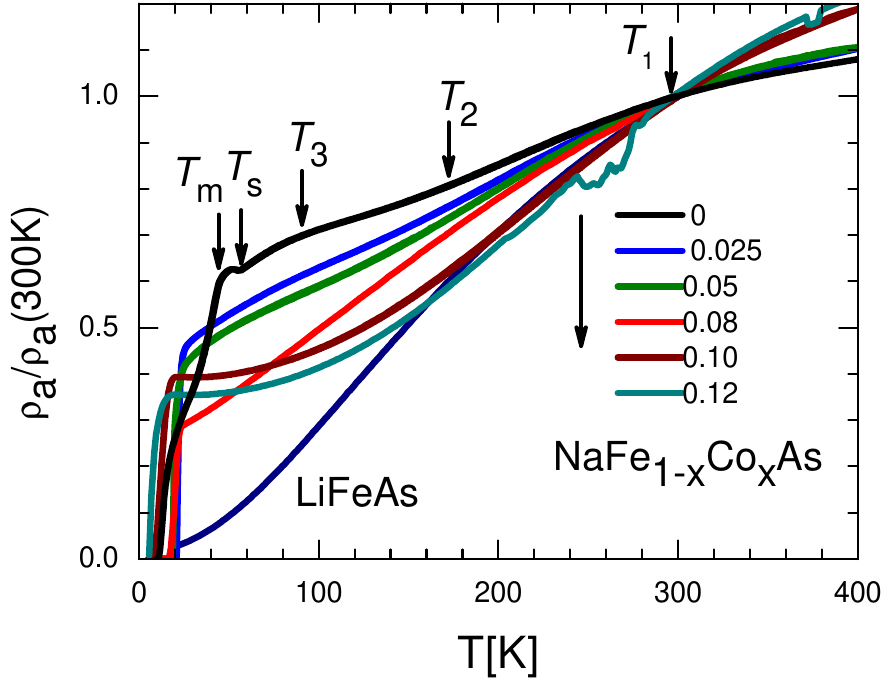}%
\caption{(Color online) Doping-evolution of the temperature dependence of the in-plane resistivity, $\rho_a / \rho _a(300K)$, for samples of NaFe$_{1-x}$Co$_x$As in fresh state after initial sample handling and contact making. For reference we show data from Ref.~\onlinecite{LiFeAsresist} for stoichiometric LiFeAs, representative of the overdoped regime. }%
\label{rhoa}%
\end{figure}

%%%%%%Figure5 Inter-plane Resistivity
\begin{figure}[tbh]%
\includegraphics[width=8cm]{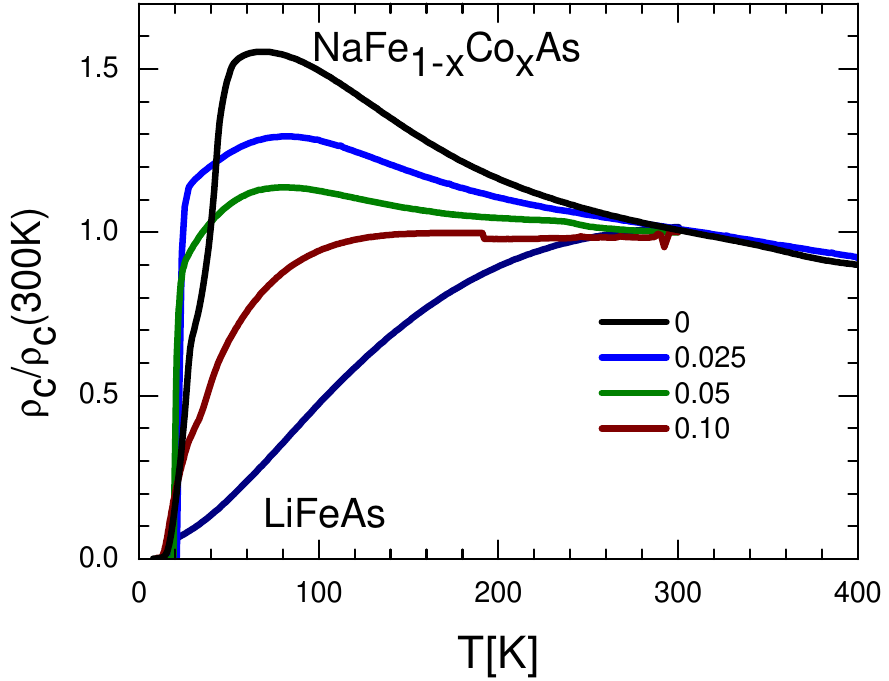}%
\caption{(Color online) Doping-evolution of the temperature dependence of the inter-plane resistivity, $\rho_c / \rho _c(300K)$, for samples of NaFe$_{1-x}$Co$_x$As in fresh state after initial sample handling and contact making. For reference we data from Ref.~\onlinecite{LiFeAsresist} for stoichiometric LiFeAs, representative of the overdoped regime. }%
\label{rhoc}%
\end{figure}

The temperature-dependent resistivity of ``fresh'' crystals of NaFe$_{1-x}$Co$_x$As is shown in Fig.~\ref{rhoa} using a normalized resistivity scale, $\rho/\rho (300K)$. The shape of $\rho _a (T)$ in the parent compound is relatively complex, with features due to split structural (at temperature $T_s$=55~K) and magnetic (at $T_m$=45~K) transitions \cite{phaseDNaFeAs,DaiNaFeAs} and slope changes at higher temperatures \cite{NaFeAs,nematicNaFeAs}. With doping the dependence transforms to very close to $T$-linear $\rho _a (T)$ in samples with $x$=0.08 and close to $T^2$ on further $x$ increase. The changes of slope at $T_1$ $\sim$300~K (increase of slope on cooling), $T_2\sim$160~K (decrease of slope on cooling), and $T_3 \sim$80~K (increase of slope on cooling) are observed in doped samples, similar to the parent compound, and are relatively insensitive to doping. The feature at $T_1$ is observed in samples with all $x$ studied. It is similar, though less pronounced, to a slope change at about the same temperature in $\rho_a (T)$ of stoichiometric LiFeAs, see Fig.~\ref{rhoa}. The features at $T_2$ and $T_3$ are observed in $\rho _a (T)$ of the samples with $x \leq$ 0.05.

The results of this study of $\rho _a (T)$ are in reasonable agreement with previous studies on single crystals \cite{nematicNaFeAs,XHChenNaFeAs}, with the difference of $x$ coming from using nominal values during sample preparation. The doping-transformation of the temperature dependent resistivity for $T$ right above $T_c$ follows general expectations for a quantum critical scenario \cite{NDL}, with $T$-linear range confined from high temperature side by slope change on approaching $T_1$. By comparison with position of maximum in the temperature dependent inter-plane resistivity $\rho _c (T)$, we assigned similar slope change feature in $\rho _a (T)$ of Ba$_{1-x}$K$_x$Fe$_2$As$_2$ to formation of pseudogap \cite{TanatarK}. The slope changes upon cooling through $T_2$ and $T_3$ in samples with $x \leq $0.05 do not have a direct analogy with Ba122 compounds. These features are observed even in samples in which long-range magnetic order and orthorhombic structural distortion are suppressed. Studies of resistivity anisotropy on detwinned single crystals of parent NaFeAs \cite{nematicNaFeAs} suggest that the feature at $T_3$ has a similar nature to nematic correlations, which is particularly strong in electron doped BaCo122 \cite{Fisher,detwinning}.

For understanding the resistivity of NaFeAs based compounds, it is important to get insight into the temperature dependence of the inter-plane resistivity component. In Fig.~\ref{rhoc} we show the doping evolution of $\rho _c (T)$  in NaFe$_{1-x}$Co$_x$As. The inter-plane resistivity of parent NaFeAs increases during cooling down to a maximum at $\sim$70~K, which is close to $T_3$ as determined from $\rho _a (T)$. The resistivity rapidly decreases below this maximum, with a notable rate increase below $T_m$. Note that contrary to $\rho _a(T)$, the inter-plane resistivity does not show an increase below $T_s$, suggesting that the carriers affected by the formation of a gap do not contribute much to the inter-plane transport. Interestingly, despite the strong difference between $\rho _a (T) $ and $\rho _c (T)$, the high temperature features are observed in both of them.

\subsection{Anisotropy of the upper critical field}

%%%%%%Figure6 Resistivity in-field
\begin{figure}[tbh]%
\includegraphics[width=8cm]{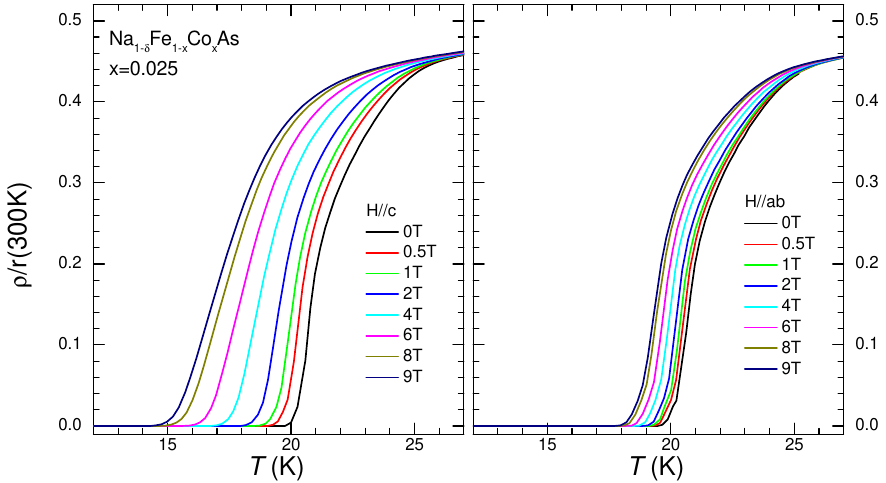}\\*
\caption{(Color online) Evolution of the temperature-dependent resistivity in the vicinity of the superconducting transition in magnetic fields applied perpendicular to ($H \parallel c$, left panel) and parallel to ($H \parallel ab$, right panel) the conducting Fe-As plane of the sample of optimally doped NaFe$_{1-x}$Co$_x$As, $x$=0.025. The resistive transition temperature, used to plot $H-T$ phase diagram shown in Fig.~\ref{Hc2T} below, was defined using midpoint criterion. }%
\label{resistivityHc2}%
\end{figure}

%%%%%%Figure7 Hc2anisotropy
\begin{figure}[tbh]%
\includegraphics[width=8cm]{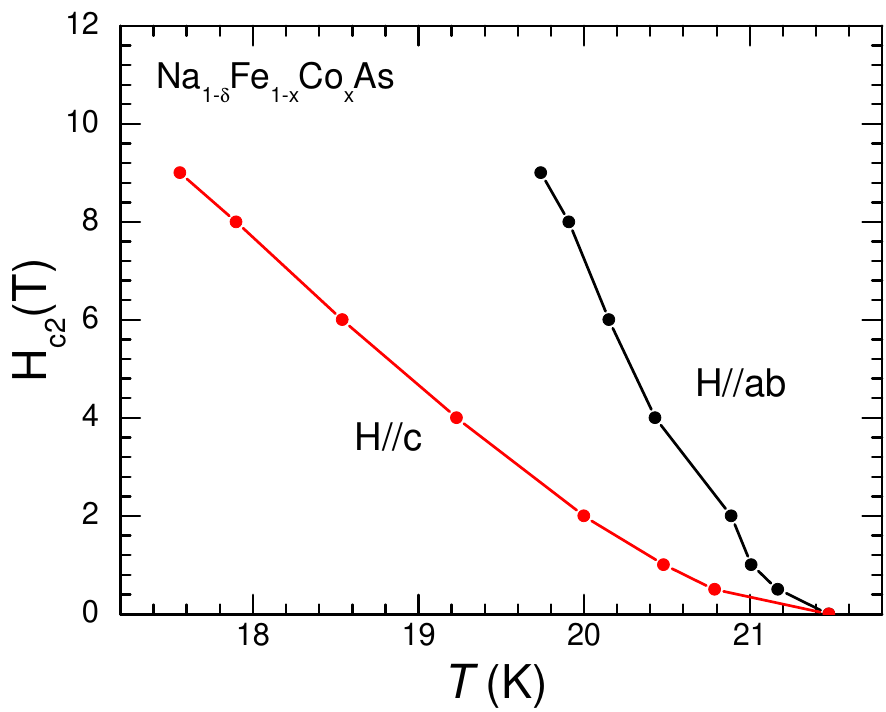}\\*
\caption{(Color online) Temperature- magnetic field phase diagram of optimally doped NaFe$_{1-x}$Co$_x$As, $x$=0.025 for orientation of magnetic field perpendicular and parallel to the conducting Fe-As plane of the crystal.
}%
\label{Hc2T}%
\end{figure}

The anisotropy of the electrical resistivity at $T_c$, $\gamma_{\rho} \equiv \frac{\rho_c (T_c)}{\rho_a (T_c)}$, is linked with the anisotropy of the upper critical field, $\gamma_H \equiv$ $\frac{H_{c2,ab}(T_c)}{H_{c2,c} (T_c)}$, with $\gamma _{\rho}= \gamma_H^2$. Because determination of the absolute values in resistivity measurements always includes uncertainty of the geometric factor and is affected by the cracks, the $H_{c2}$ anisotropy measurements provide an alternative way to evaluate resistivity anisotropy \cite{anisotropy}. In Fig.~\ref{resistivityHc2} we zoom the superconducting transition in in-plane resistivity measurements $\rho _a (T)$ for a sample with doping level close to optimal, $x$=0.025. The same sample was remounted on a plastic cube with the magnetic field was in the $H \parallel c$ (left panel) and $H \parallel ab$ (right panel) configurations.

We used the resistive transition midpoint to determine $H_{c2}(T)$ anisotropy as shown in Fig.~\ref{Hc2T}. Close to $T_c$ the anisotropy $\gamma _H$=2.25 $\pm 0.1$ for the sample with $x$=0.025. Similar value with $\gamma _H$=2.35 $\pm 0.1$  was obtained in sample with $x$=0.08. These measurements suggest a resistivity anisotropy of about 5 at $T_c$. Considering that $\gamma _{\rho}(T_c) \sim 2 \gamma _{\rho} (300~K)$ (see Figs.~\ref{rhoa} and \ref{rhoc}) we expect a negligible anisotropy of 2 to 3 at room temperature. The direct resistivity measurements, with $\rho _a (300K)$=400 to 500 $\mu \Omega$cm and  $\rho _c (300K)$=2000 to 3000 $\mu \Omega$cm, suggest an anisotropy of 4 to 8. The origin of this factor of about two discrepancy remains unclear at the moment.

\section{Discussion}

\subsection{ Slope-change features in the temperature-dependent resistivity }

%%%%%%Figure8 rhoa vs rhoc parent
\begin{figure}[tbh]%
\includegraphics[width=8cm]{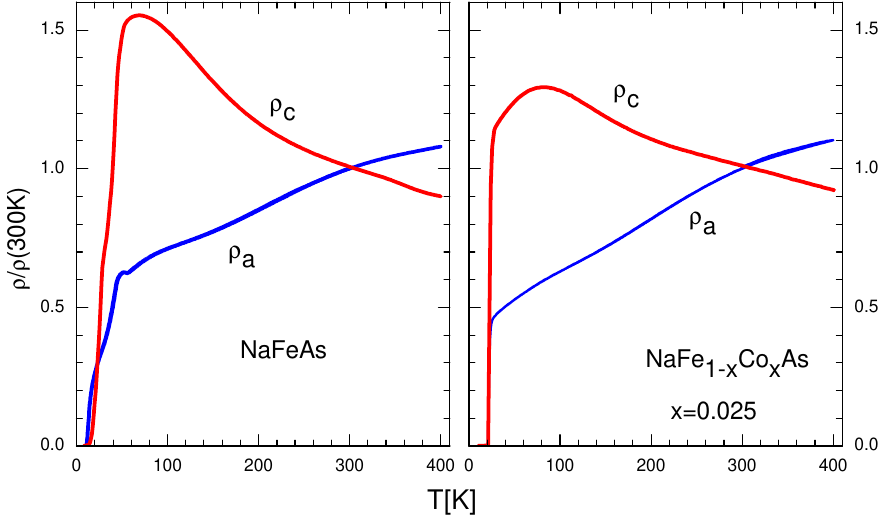}\\*
\caption{(Color online) Comparison of the temperature- dependent in-plane and inter-plane resistivity of parent NaFeAs (left panel) and a sample with $x$=0.025 (right panel). Resistivity data are plotted vs. normalized scale, $\rho (T) / \rho (300K)$.
}%
\label{avsc}%
\end{figure}

%%%%%%Figure9 rhoa Co vs environmental
\begin{figure}[tbh]%
\includegraphics[width=8cm]{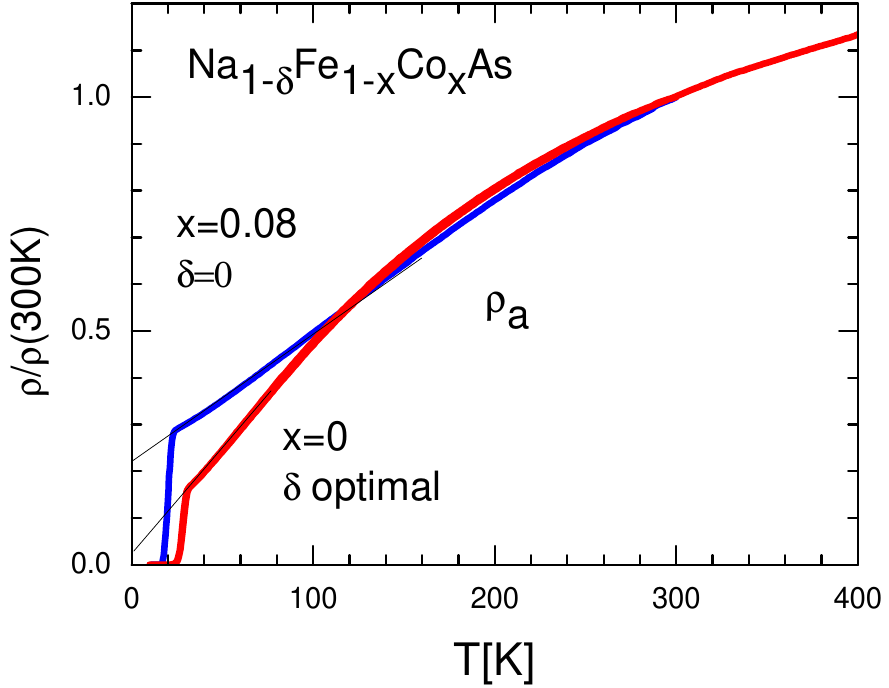}\\*
\caption{(Color online) Comparison of the temperature- dependent in-plane resistivity in samples of NaFeAs doped to the highest $T_c$ by environmental reaction with Apiezon N-grease and with Co-doping, $x$=0.05. Resistivity data are plotted vs. normalized scale, $\rho (T) / \rho (300K)$. Lines show linear extrapolation of the curves to $T \to $0, revealing difference in residual resistivity of two concentrations of samples.
}%
\label{Covsenvironmental}%
\end{figure}

As can be seen from direct comparison of in-plane and inter-plane resistivity in the parent and slightly doped $x$=0.025 compositions, Fig.~\ref{avsc}, features in $\rho_a (T)$ find counterparts in $\rho _c(T)$. For example, a slope decrease in generally metallic $\rho_a (T)$ below $T_2 \sim$160~K is seen as slope increase in generally activated (increasing on cooling) $\rho_c (T)$. This similarity found in two very different dependences suggests that the activation of carriers over a partial gap, rather than change of scattering, is responsible for the feature. Partial (nematic) order \cite{nematicNaFeAs}, which happens above the structural transition at $T_3$, changes $\rho_c (T)$ from insulating to metallic, while the magnetic order below $T_m$ causes a dramatic decrease of resistivity in both directions of charge flow. The decrease is especially strong in the parent compound in which the residual resistivity ratio (RRR), $\rho (300K)/ \rho (T_c)$, is a factor of two higher than in $x$=0.025. These observations suggests that magnetic scattering plays an important role in resistivity at $T>T_3$, and that when inelastic scattering is dominant (as in the parent compound) taming down of magnetic fluctuations reveals intrinsically very low residual resistivity.

\subsection{Residual resistivity}

Observation of a much higher RRR in non-doped materials agrees with studies in Ba122 compounds, though in the latter, the direct comparison is not so simple. In the case of NaFeAs based materials we can compare RRR of the samples, brought to optimal doping using two different doping, electron with Co substitution of Fe and environmental, on interaction with the environment. In Fig.~\ref{Covsenvironmental} we compare $T$-dependent resistivity in two concentrations of samples, extrapolating curves linearly from $T_c$ to $T =0$.
The RRR ratio decreases from more than 20 in environmentally doped samples to about 4 in Co-doped samples. Taking that resistivity at room temperature does not change from about 400 $\mu \Omega$cm , this suggests that the residual resistivity induced by the $x$=0.025 substitution of Fe atoms with Co is on the order of 100 $\mu \Omega$cm, comparable to BaCo122 \cite{pseudogap}. This is almost a factor of five higher than the value found in very disordered samples, doped with environmental reaction, which extrapolates to $\rho_0 \approx$20 $\mu \Omega$cm.

\section{Conclusions}

In conclusion, we find that the complicated shape of the temperature dependent inter-plane resistivity of both parent NaFeAs and Co - doped NaFe$_{1-x}$Co$_x$As shows the same anomalies as in-plane resistivity. This is particularly interesting considering the fact that inter-plane transport is clearly thermally activated, while the in - plane resistivity follows metallic decrease on cooling. This finding suggests that the observed features are not caused by a particular type of scattering process and most likely are determined by the variation in the carrier density. Such behavior strongly supports the idea that these features are caused by the thermal activation of charge carriers over the pseudogap in the electronic spectrum. This conclusion suggests that the pseudogap is a common feature of both NaFeAs - based materials and BaFe$_2$As$_2$ - derived compounds \cite{pseudogap,pseudogap2,BasovPseudogap}.

%%%%%%%%%%%%%%%%%%%%%%%%%%%% ACKNOWLEDGMENTS
\section{Acknowledgements}

We thank Seyeon Park for her help with dipper measurements.
Work at the Ames Laboratory was supported by the Department of Energy-Basic Energy Sciences under Contract No. DE-AC02-07CH11358. The single crystal growth effort at UT is supported by U.S. DOE BES under Grant No. DE-FG02-05ER46202 (P.D.).

%%%%%%%%%%%%%%%%%%%%%%%%%%%% BIBLIOGRAPHY

\end{document}